\documentclass[12pt]{revtex4}
\usepackage{dcolumn}
\usepackage{graphicx}
\usepackage{amsmath}
\usepackage{amsfonts}
\usepackage{amssymb}
\usepackage{psfrag}
\usepackage{wrapfig}
\usepackage{subfigure}
\usepackage{makeidx}
\usepackage{bm}
\usepackage{epsf}
\usepackage{hyperref}

\begin{document}

\title{Slowly rotating charged black holes in anti-de Sitter third order Lovelock gravity}

\author{ Rui-Hong Yue$^{1}$\footnote{ Email:yueruihong@nbu.edu.cn}, De-Cheng Zou$^{2}$,
Tian-Yi Yu$^{1}$, Peng Li$^{3}$ and Zhan-Ying Yang$^{2}$}

\affiliation{ $^{1}$Faculty of Science, Ningbo University, Ningbo 315211, China\\
$^{2}$Department of Physics, Northwest University, Xi'an, 710069, China
\\$^{3}$Institute of Modern Physics, Northwest University,
Xi'an, 710069, China}

\begin{abstract}
\indent

In this paper, we study slowly rotating black hole solutions in Lovelock gravity ($n=3$).
These exact slowly rotating black hole solutions are obtained in uncharged and charged cases,
respectively. Up to the linear order of the rotating parameter $a$, the mass,
Hawking temperature and entropy of the uncharged black holes get no corrections from rotation.
In charged case, we compute magnetic dipole moment and gyromagnetic ratio of the black holes.
It is shown that the gyromagnetic ratio keeps invariant after introducing the Gauss-Bonnet
and third order Lovelock interactions.
\end{abstract}



\maketitle

\section{Introduction}
\indent

It is believed that Einstein's gravity is a low-energy limit of a quantum theory of gravity.
Considering the fundamental nature of quantum gravity, there should be a low-energy effective
action which describes gravity at the classical level \cite{Dehghani:2003cu}. In addition to
Einstein-Hilbert action, this effective action also involves higher derivative terms, and these
higher derivative terms can be seen in the renormalization of quantum field theory in curved
spacetimes \cite{Birrell:1982ix}, or in the construction of the low-energy effective action of
string \cite{Lust:1989tj}. In the AdS/CFT correspondence, the higher derivative terms can be
regarded as the corrections of large N expansion in the dual conformal field theory. In general,
the higher powers of curvature can give rise to a fourth or even higher order differential
equation for the metric, and it will introduce ghosts and violate unitarity. So, the higher
derivative terms may be a source of inconsistencies. However, Zwiebach and Zumino \cite{Zwiebach:1985uq}
found that the ghosts can be avoided if the higher derivative terms only consist of the
dimensional continuations of the Euler densities, leading to second order field equations for
the metric \cite{Ge:2009ac}. This higher derivative  theory is so-called Lovelock
gravity \cite{Lovelock:1971yv}, and the equations of motion contain the most symmetric
conserved tensor with no more than second derivative of the metric. In this paper, we
indulge ourselves with the first four terms of the Lovelock gravity, corresponding to
the cosmological constant, Einstein term, Gauss-Bonnet and third order Lovelock
terms respectively. So far, the exact static and spherically symmetric black hole
solutions in third order Lovelock gravity were first found in \cite{Dehghani:2009zzb},
and the thermodynamics have
been investigated in \cite{Ge:2009ac, Dehghani:2009zzb, Dehghani:2005zzb}.

On the other hand, a great many attentions have been focused on these static and spherically
symmetric black hole solutions by the effect of rotation. In the AdS/CFT correspondence,
the rotating black holes in AdS space are dual to certain CFTs in a rotating
space \cite{Hawking:1998kw}, while charged ones are dual to CFTs with chemical
potential \cite{Cvetic:1999rb}. In general relativity, the higher dimensional rotating black
holes have been recently studied and some exact analytical solutions of Einstein's equation
were found in \cite{Gibbons:2004uw, Aliev:2007qi}.

Since the equations of motion of Lovelock gravity are highly nonlinear, it is rather difficult
to obtain the explicit rotating black hole solutions. A new method is needed. In order to find
rotating black hole solutions in the presence of dilaton coupling electromagnetic field in
Einstein(-Maxwell) theory, Horne and Horowitz \cite{Horne:1992zy} first developed a simple
perturbative method that a small angular momentum as a perturbation was introduced into a
non-rotating system, and obtained slowly rotating dilaton black hole solutions. Until now,
this approach has been extensively discussed in general relativity \cite{Sheykhi:2008rm}.
Taking advantage of this crucial tool, Kim and Cai \cite{Kim:2007iw}
studied slowly rotating black hole solutions with one nonvanishing angular momentum in
the Gauss-Bonnet gravity, here the rotating parameter $a$ appears as a small quantity.
Recently, some numerical results about the existence of five-dimensional rotating
Gauss-Bonnet black holes with angular momenta of the same magnitude have been presented
in \cite{Brihaye:2008kh}. In addition, it is worth to mention that some rotating
black brane solutions have been investigated in the second (Gauss-Bonnet) and
third order Lovelock gravity \cite{Dehghani2006cu}. Nevertheless, these solutions
are essentially obtained by a Lorentz boost from corresponding static ones.
They are equivalent to static ones locally, although not equivalent globally.
In this paper, we will analyze slowly rotating black hole solutions in third order Lovelock
gravity. Following the Horne and Horowitz's perturbative method, a small rotating parameter
$a$ as a perturbation into the metric will be introduced.
The slowly rotating black hole solutions will be studied in uncharged and charged cases,
and then we analyze some physical properties of these black holes.

The outline of this paper is as follows. In section \ref{2s}, we review the $(n=3)$
Lovelock gravity, and derive the equations of gravitation and electromagnetic fields.
Then, we explore slowly rotating uncharged black holes and obtain the slowly rotating
black hole solution $f(r)$ and expression for function $p(r)$
by putting a new form metric into these equations. Moveover,
we discuss some related physical properties of the black holes. In section \ref{3s},
we set about learning slowly rotating black holes in charged case. Section \ref{4s}
is devoted to conclusions and discussions.

\section{Slowly Rotating Black Holes in Uncharged Case\label{2s}}
\subsection{Action and Black Hole Solutions}
\indent

The action of third order Lovelock gravity in the presence of electromagnetic field can
be written as
\begin{eqnarray}
{\cal I}=\frac{1}{16\pi G}\int{d^{D}x\sqrt{-g}(-2\Lambda+\alpha_{1}{\cal L}_{1}
+\alpha_{2}{\cal L}_{2}+\alpha_{3}{\cal L}_{3}-4\pi G F_{\mu\nu}F^{\mu\nu})}\label{1a},
\end{eqnarray}
where $\alpha_{i}$ is the $i$-th order Lovelock coefficients,
$F_{\mu\nu}=\partial_{\mu}A_{\nu} -\partial_{\nu}A_{\mu}$ is
electromagnetic field tensor with a vector potential $A_{\mu}$. The Einstein term ${\cal L}_1$
equals to $R$, and the second order Lovelock(Gauss-Bonnet) term ${\cal L}_2$ is
$R_{\mu\nu\sigma\kappa}R^{\mu\nu\sigma\kappa}-4R_{\mu\nu}R^{\mu\nu}+R^2$.
${\cal L}_{3}$ measures the third order Lovelock term which described as
\begin{eqnarray}
&&2R^{\mu\nu\sigma\kappa}R_{\sigma\kappa\rho\tau}R^{\rho\tau}_{~~\mu\nu}
+8R^{\mu\nu}_{~~\sigma\rho}R^{\sigma\kappa}_{~~\nu\tau}R^{\rho\tau}_{~~\mu\kappa}
+24R^{\mu\nu\sigma\kappa}R_{\sigma\kappa\nu\rho}R^{\rho}_{\mu}+\nonumber\\
&&3RR^{\mu\nu\sigma\kappa}R_{\mu\nu\sigma\kappa}
+24R^{\mu\nu\sigma\kappa}R_{\sigma\mu}R_{\kappa\nu}
+16R^{\mu\nu}R_{\nu\sigma}R^{\sigma}_{~\mu}-12RR^{\mu\nu}R_{\mu\nu}+R^3.\label{2a}
\end{eqnarray}

Varying the action with respect to the metric tensor $g_{\mu\nu}$ and electromagnetic tensor
field $F_{\mu\nu}$, the equations for gravitation and electromagnetic fields are
\begin{eqnarray}
\Lambda
g_{\mu\nu}+\alpha_{1}G^{(1)}_{\mu\nu}+\alpha_{2}G^{(2)}_{\mu\nu}+\alpha_{3}G^{(3)}_{\mu\nu}
=8\pi GT_{\mu\nu}, \label{3a}
\end{eqnarray}
\begin{eqnarray}
\partial_{\mu}(\sqrt{-g}F^{\mu\nu})=0.\label{4a}
\end{eqnarray}
Here $T_{\mu\nu}=F_{\mu\alpha}F_{\nu}^{~\alpha}
-\frac{1}{4}g_{\mu\nu}F_{\alpha\beta}F^{\alpha\beta}$
is the energy-momentum tensor of electromagnetic field, $G^{(1)}_{\mu\nu}=R_{\mu\nu}
-\frac{1}{2}Rg_{\mu\nu}$ is Einstein tensor, and $G^{(2)}_{\mu\nu}$ and $G^{(3)}_{\mu\nu}$ are
the second order Lovelock(Gauss-Bonnet) and third order Lovelock  tensors respectively:
\begin{eqnarray}
G^{(2)}_{\mu\nu}=2(R_{\mu\sigma\kappa\tau}R_{\nu}^{~\sigma\kappa\tau}
-2R_{\mu\rho\nu\sigma}R^{\rho\sigma}
-2R_{\mu\sigma}R^{\sigma}_{~\nu}+RR_{\mu\nu})-\frac{1}{2}L_{2}g_{\mu\nu} ,\nonumber
\end{eqnarray}
\begin{eqnarray}
G^{(3)}_{\mu\nu}&=&3R_{\mu\nu}R^2-12RR_{\mu}^{~\sigma}R_{\sigma\nu}
-12R_{\mu\nu}R_{\alpha\beta}R^{\alpha\beta}
+24R_{\mu}^{~\alpha}R_{\alpha}^{~\beta}R_{\beta\nu}-24R_{\mu}^{~\alpha}R^{\beta\sigma}R_{\alpha
\beta\sigma\nu}\nonumber\\&+&3R_{\mu\nu}R_{\alpha\beta\sigma\kappa}R^{\alpha\beta\sigma\kappa}
-12R_{\mu\alpha}R_{\nu\beta\sigma\kappa}R^{\alpha\beta\sigma\kappa}
-12RR_{\mu\sigma\nu\kappa}R^{\sigma\kappa}
+6RR_{\mu\alpha\beta\sigma}R_{\nu}^{~\alpha\beta\sigma}\nonumber\\
&+&24R_{\mu\alpha\nu\beta}R_{\sigma}^{~\alpha}R^{\sigma\beta}+24R_{\mu\alpha\beta\sigma}R_{
\nu}^{~\beta}R^{\alpha\sigma}+24R_{\mu\alpha\nu\beta}R_{\sigma\kappa}R^{\alpha\sigma\beta\kappa}
-12R_{\mu\alpha\beta\sigma}R^{\kappa\alpha\beta\sigma}R_{\kappa\nu}\nonumber\\
&-&12R_{\mu\alpha\beta\sigma}R^{\alpha\kappa}R_{\nu\kappa}^{~~\beta\sigma}
+24R_{\mu}^{~\alpha\beta\sigma}R_{\beta}^{~\kappa}R_{\sigma\kappa\nu\alpha}
-12R_{\mu\alpha\nu\beta}R^{\alpha}_{~\sigma\kappa\rho}R^{\beta\sigma\kappa\rho}\nonumber\\
&-&6R_{\mu}^{~\alpha\beta\sigma}R_{\beta\sigma}^{~~\kappa\rho}R_{\kappa\rho\alpha\nu}
-24R_{\mu\alpha}^{~~\beta\sigma}R_{\beta\rho\nu\lambda}R_{\sigma}^{~\lambda\alpha\rho}
-\frac{1}{2}L_{3}g_{\mu\nu} .\nonumber
\end{eqnarray}

Usually,  the action Eq.~(\ref{1a}) is supplemented with surface terms (a Gibbons-Hawking surface term)
whose variation will  cancel the extra normal derivative term in deriving the equation of motion Eq.~(\ref{3a}).
However, these surface terms is not necessary in our discussion and will be neglected.
Note that for third order Lovelock gravity, the nontrivial third term requires the dimension(D) of
spacetime satisfying $D\geq7$.

The metric of slowly rotating spacetime can be written as \cite{Kim:2007iw}
\begin{eqnarray}
ds^2=-f(r)dt^2+\frac{1}{f(r)}dr^2+\sum^{D}_{i=j=3}r^2h_{ij}dx^idx^j-2ar^2p(r)h_{44}dtd\phi,\label{5a}
\end{eqnarray}
where $h_{ij}dx^idx^j$ represents the metric of a $(D-2)$-dimensional hyper-surface with constant
curvature scalar $(D-2)(D-3)k$ and volume $\Sigma_{k}$, here k is a constant. Without loss of generality,
one can take $k=0$ or $\pm1$. When $k=1$, one has $h_{ij}dx^idx^j=d\theta^2+\sin^2\theta d\phi^2 +\cos^2\theta
d\Omega^2_{D-4}$ and $h_{44}=\sin^2\theta$; when $k=0$, $h_{ij}dx^idx^j=d\theta^2 +d\phi^2+dx^{2}_{D-4}$
and $h_{44}=1$; when $k=-1$, $h_{ij}dx^idx^j=d\theta^2+\sinh^2\theta d\phi^2 +\cosh^2\theta d\Omega^2_{D-4}$
and $h_{44}=\sinh^2\theta$, where $dx^2_{D-4}$ is the line element of a $(D-4)$-dimensional Ricci flat
Euclidian surface. While $d\Omega^2_{D-4}$ denotes the line element of a $(D-4)$-dimensional unit sphere.

For the convenience future, we introduce new parameters $\tilde{\alpha}_i$
\begin{eqnarray}
&&\tilde{\alpha}_0=\frac{2\Lambda}{(D-1)(D-2)},\qquad
\tilde{\alpha}_{i}=\alpha_i\prod^{2i-2}_{l=1}(D-2-l),\quad(i=1, 2, 3).\label{6a}
\end{eqnarray}
Firstly, we consider the case without charge; namely $T_{\mu\nu}=0$. Solving Eq.~(\ref{3a})
for the metric given in Eq.~(\ref{5a}) and discarding any terms involving $a^2$ or
higher powers, we find that the $rr$-component of the equations of motion
\begin{eqnarray}
0&=&(D-7)\tilde{\alpha}_3(f(r)-k)^3-(D-5)\tilde{\alpha}_2(f(r)-k)^2r^2
+(D-3)\tilde{\alpha}_1(f(r)-k)r^4\nonumber\\
&+&[3\tilde{\alpha}_3r(f(r)-k)^2+2\tilde{\alpha}_2(f(r)-k)r^3+\tilde{\alpha}_1 r^5]f'(r)
+(D-1)\tilde{\alpha}_0 r^6,\label{7a}
\end{eqnarray}
where a prime denotes the derivative with respect to $r$. We notice that the angular momentum
parameter $a$ does not appear in the $rr$-component. Thus, the slowly rotating
black hole solutions $f(r)$ is identical to the static one in form. In Eq.~(\ref{7a}),
there exist one real and two complex solutions $f(r)$. Here, we only take the real one.
This general solution $f(r)$ for D-dimensional slowly rotating black hole is
\begin{eqnarray}
f(r)=k+\frac{r^2}{3\tilde{\alpha}_3}\Big[\tilde{\alpha}_2+\sqrt[3]{\sqrt{\gamma
+\kappa^2(r)}+\kappa(r)}
-\sqrt[3]{\sqrt{\gamma+\kappa^2(r)}-\kappa(r)}\Big],\label{8a}
\end{eqnarray}
where
\begin{eqnarray}
\gamma=(3\tilde{\alpha}_1\tilde{\alpha}_3-{\tilde{\alpha}_2}^2)^3,\quad
\kappa(r)={\tilde{\alpha}_2}^3-\frac{9\tilde{\alpha}_1\tilde{\alpha}_2\tilde{\alpha}_3}{2}
-\frac{27{\tilde{\alpha}_3}^2}{2}[\tilde{\alpha}_0+\frac{16\pi
GM}{(D-2)\Sigma_kr^{D-1}}].\label{9a}
\end{eqnarray}
The integral constant $M$ is the gravitational mass. Hereafter, for simplicity, we
take notation $m=\frac{16\pi GM}{(D-2)\Sigma_k}$. It is easy to find that the solution is
asymptotically flat for $\Lambda=0$, AdS for negative value of $\Lambda$ and dS for positive value
of $\Lambda$. We discuss the case of asymptotically AdS solutions in this paper.
Thus, putting $\tilde{\alpha}_0=-1/l^2$ in Eq.~(\ref{9a}), we obtain
\begin{eqnarray}
\kappa(r)&=&{\tilde{\alpha}_2}^3-\frac{9\tilde{\alpha}_1\tilde{\alpha}_2\tilde{\alpha}_3}{2}
-\frac{27{\tilde{\alpha}_3}^2}{2}\Big[-\frac{1}{l^2}+\frac{m}{r^{D-1}}],\nonumber\\
\varphi&=&-\frac{1}{3\tilde{\alpha}_3}\Big[\tilde{\alpha}_2+\sqrt[3]{\sqrt{\gamma+\kappa^2(r)}+\kappa(r)}
-\sqrt[3]{\sqrt{\gamma+\kappa^2(r)}-\kappa(r)}\Big],\label{10a}
\end{eqnarray}
where $\varphi=(k-f(r))/{r^2}$.

Meanwhile, there exists off-diagonal $t\phi$-component of equations of motion, which is
concerned with function $p(r)$. A tedious computation leads to a following equation
\begin{eqnarray}
\frac{A(r)}{2}p''(r)+\frac{[3A(r)+(D-3)B(r)]}{2r}p'(r)=0,\label{11a}
\end{eqnarray}
where
\begin{eqnarray}
A(r)&=&\tilde{\alpha}_1+2\tilde{\alpha}_2\varphi+3\tilde{\alpha}_3\varphi^2,\nonumber\\
B(r)&=&\tilde{\alpha}_1+2\tilde{\alpha}_2\varphi+3\tilde{\alpha}_3\varphi^2
+\frac{2r\tilde{\alpha}_2\varphi'}{D-3}+\frac{6r\tilde{\alpha}_3\varphi\varphi'}{D-3}.\label{12a}
\end{eqnarray}
It can be changed into a closed form
\begin{eqnarray}
[\log p'(r)]'&=&-[\frac{D}{r}+\frac{(\tilde{\alpha}_1+2\tilde{\alpha}_2\varphi+3\tilde{\alpha}_3
\varphi^2)'}{\tilde{\alpha}_1+2\tilde{\alpha}_2\varphi+3\tilde{\alpha}_3\varphi^2}]\nonumber\\
&=&-[\log(r^D(
\tilde{\alpha}_1+2\tilde{\alpha}_2\varphi+3\tilde{\alpha}_3\varphi^2))]'.\label{13a}
\end{eqnarray}
Therefore, the formal expression for function $p(r)$ in third order Lovelock gravity is given by
\begin{eqnarray}
p(r)=\int{\frac{C_2dr}{r^D(\tilde{\alpha}_1+2\tilde{\alpha}_2\varphi+3\tilde{\alpha}_3\varphi^2)}}
+C_1,\label{14a}
\end{eqnarray}
where the $C_{1}$ and $C_2$ are two integration constants.

Note that the exact static and spherically symmetric black hole solutions of third
order Lovelock gravity have been found by working directly in the action \cite{Dehghani:2009zzb}.
Here, we adopt the same approach. We substitute the metric Eq.~(\ref{5a}) into the
action Eq.~(\ref{1a}), and then it reduces to
\begin{eqnarray}
{\cal I}=\frac{(D-2)\Omega_{D-2}}{16\pi G}\int dtdr[\frac{r^{D-1}}{l^2}+r^{D-1}\varphi(\tilde{\alpha}_1+\tilde{\alpha}_2\varphi
+\tilde{\alpha}_3\varphi^2)]',\label{15a}
\end{eqnarray}
where a prime denotes derivative to $r$. Clearly, the Eq.~(\ref{15a}) is similar in form to
the static black hole solution \cite{Dehghani:2009zzb}. By varying the action with respect to $f(r)$,
one obtains the equation of motion
\begin{eqnarray}
[\frac{r^{D-1}}{l^2}+r^{D-1}\varphi(\tilde{\alpha}_1+\tilde{\alpha}_2\varphi
+\tilde{\alpha}_3\varphi^2)]'=0.\label{16a}
\end{eqnarray}
Therefore, $\varphi$ is determined by solving for the real roots of the following 3th-order
polynomial equation
\begin{eqnarray}
\tilde{\alpha}_1\varphi+\tilde{\alpha}_2\varphi^2
+\tilde{\alpha}_3\varphi^3=\frac{m}{r^{D-1}}-\frac{1}{l^2}.\label{17a}
\end{eqnarray}
We can easily verify by drawing a parallel between the Eqs.(\ref{14a})(\ref{17a})
\begin{eqnarray}
p(r)=\frac{C_2\varphi}{m(1-D)}+C_1.\label{18a}
\end{eqnarray}
Let the constants $C_2=m(D-1)$ and $C_1=0$, the function $p(r)$ can be written as
\begin{eqnarray}
p(r)=-\varphi=\frac{1}{3\tilde{\alpha}_3}\Big[\tilde{\alpha}_2+\sqrt[3]{\sqrt{\gamma
+\kappa^2(r)}+\kappa(r)}-\sqrt[3]{\sqrt{\gamma+\kappa^2(r)}-\kappa(r)}\Big].\label{19a}
\end{eqnarray}

\subsection{Physical properties}
\indent

As shown in Eq.~(\ref{8a}), the slowly rotating black hole solution $f(r)$ is independent of $a$.
Though most interesting physical properties also depend only on $a^2$, one can still extract
some useful information from it. Based on discussions in the last subsection,
we will investigate physical properties of slowly rotating black holes in this subsection.

According to the solution $f(r)$, the gravitational mass of the solution can be expressed as
\begin{eqnarray}
M=\frac{(D-2)\Sigma_{k}r_{+}^{D-7}}{16\pi
G}(r_{+}^6/l^2+k\tilde{\alpha}_1 r_{+}^4+k^2\tilde{\alpha}_2 r_{+}^2+k^3\tilde{\alpha}_3)\label{20a}
\end{eqnarray}
and the Hawking temperature of the black hole is
\begin{eqnarray}
T&=&\frac{f'(r_{+})}{4\pi}\nonumber\\
&=&\frac{(D-1)r_{+}^6/l^2+(D-3)k\tilde{\alpha}_1 r_{+}^4
+(D-5)k^2\tilde{\alpha}_2 r_{+}^2
+(D-7)k^3\tilde{\alpha}_3}{4\pi r_{+}(\tilde{\alpha}_1r_{+}^4+2k\tilde{\alpha}_2 r_{+}^2+3k^2\tilde{\alpha}_3)}.\label{21a}
\end{eqnarray}
Thus, the angular momentum of the black hole
\begin{eqnarray}
J=\frac{2aM}{D-2}=\frac{a\Sigma_{k}r_{+}^{D-7}}{8\pi G}(r_{+}^6/l^6+k\tilde{\alpha}_1 r_{+}^4
+k^2\tilde{\alpha}_2 r_{+}^2+k^3\tilde{\alpha}_3).\label{22a}
\end{eqnarray}

Another important thermodynamic quantity is black hole entropy. Usually, the entropy of
black hole satisfies the so-called area law of entropy which states that the black hole entropy equals
to one-quarter of the horizon area \cite{Gibbons:1977mu, Hawking:1974rv}. It applies to all kinds of
black holes and black strings of Einstein gravity \cite{Hunter:1998qe}. However, in higher derivative
gravity, the area law of the entropy is not satisfied in general \cite{Jacobson:1993xs}. Since black
hole can be regard as a thermodynamic system, it obeys the first law of thermodynamics $dM=TdS+\omega_{H}dJ$.
Through the angular velocity $\omega_{H}$, one can get the entropy of black hole.

For the slowly rotating solution, the stationarity and  rotational symmetry metric Eq.~(\ref{5a})
admits two commuting Killing vector fields
\begin{eqnarray}
\xi_{(t)}=\frac{\partial}{\partial t}, \quad \xi_{\phi}=\frac{\partial}{\partial \phi}\label{23a}.
\end{eqnarray}
The various scalar products of these Killing vectors can be expressed through the metric components
as follows
\begin{eqnarray}
\xi_{(t)}\cdot\xi_{(t)}&=&g_{tt}=-f(r),\nonumber\\
\xi_{(t)}\cdot\xi_{(\phi)}&=&g_{t\phi}=-ar^2p(r)h_{44},\nonumber\\
\xi_{(\phi)}\cdot\xi_{(\phi)}&=&g_{\phi\phi}=r^2h_{44}.\nonumber
\end{eqnarray}

To examine further properties of the slowly rotating black holes, as well as physical processes near
such a black hole, we introduce a family of locally non-rotating observers. The coordinate angular
velocity for these  observers that move on orbits with constant $r$ and $\theta$ and with a
four-velocity $u^{\mu}$ such that $u\cdot\xi_(\phi)=0$ is given by \cite{Aliev:2007qi, Kim:2007iw}
\begin{eqnarray}
\Omega&=&-\frac{g_{t\phi}}{g_{\phi\phi}}=ap(r)\nonumber\\
&=&\frac{a}{3\tilde{\alpha}_3}\Big[\tilde{\alpha}_2
+\sqrt[3]{\sqrt{\gamma+\kappa^2(r)}+\kappa(r)}
-\sqrt[3]{\sqrt{\gamma+\kappa^2(r)}-\kappa(r)}\Big].\label{24a}
\end{eqnarray}
In contrast to the case of an ordinary kerr black hole in asymptotically flat spacetime, the angular
velocity does not vanish at spatial infinity
\begin{eqnarray}
\Omega_\infty=\frac{a}{3\tilde{\alpha}_3}\Big[\tilde{\alpha}_2
+\sqrt[3]{\sqrt{\gamma+\tilde{\kappa}^2(r)}+\tilde{\kappa}(r)}
-\sqrt[3]{\sqrt{\gamma+\tilde{\kappa}^2(r)}-\tilde{\kappa}(r)}\Big]
=a\Delta.\label{25a}
\end{eqnarray}
where $\kappa(r)={\tilde{\alpha}_2}^3-\frac{9\tilde{\alpha}_1\tilde{\alpha}_2\tilde{\alpha}_3}{2}
+\frac{27{\tilde{\alpha}_3}^2}{2l^2}$ and $\Delta$ is a constant.

When approaching the black hole horizon, the angular velocity turns to be
$\Omega_H=ap(r_+)=-a\varphi(r_+)=-\frac{ak}{r_+^2}$. This $\Omega_H$ can be thought as
the angular velocity of the black hole. The relative angular velocity  with respect to
a frame static at infinity is defined by
\begin{eqnarray}
\omega_H=\Omega_H-\Omega_\infty=-a(\frac{k}{r_+^2}+\Delta).\label{26a}
\end{eqnarray}
Therefore, we get the entropy of slowly rotating black hole up to the linear order of the rotating
parameter $a$
\begin{eqnarray}
S=\frac{\Sigma_{k}}{4G}r_{+}^{D-2}[\tilde{\alpha}_1+\frac{2(D-2)k\tilde{\alpha}_2}{(D-4)r_{+}^2}
+\frac{3(D-2)k^2\tilde{\alpha}_3}{(D-6)r_{+}^4}],\label{27a}
\end{eqnarray}
which recovers the results in \cite{Dehghani:2009zzb}.

\section{Slowly Rotating Black Holes in Charged Case\label{3s}}
\indent

In this section, we consider slowly rotating black hole solution with charge.
In charged case, the situation is dramatically altered. Since the black hole rotates
along the direction $\phi$,
it will generate a magnetic field. Considering this effect, the gauge potential can be chosen
\begin{eqnarray}
A_{\mu}dx^{\mu}=A_{t}dt+A_{\phi}d\phi.\label{28a}
\end{eqnarray}
Here we assume $A_{\phi}=-aQc(r)h_{44}$. As a result, the electro-magnetic field associated with
the solution are
\begin{eqnarray}
F_{tr}=-A_t',\quad F_{r\phi}=-a Q c'(r)h_{44},
\quad F_{\theta\phi}=-a Q c(r)\frac{d(h_{44})}{d\theta}.\label{29a}
\end{eqnarray}
where $Q$, an integration constant, is the electric charge of the black hole and a prime denotes
the derivative with respect to $r$. Form $t$-component of electromagnetic field
equation $\partial_{\mu}(\sqrt{-g}F^{\mu\nu})=0$, one can find $F_{tr}=\frac{Q}{4\pi r^{D-2}}$,
which is the same as the static form. Unlike the static case, there exist the $\phi$-component of
the electromagnetic field equation, and then the equation for function $c(r)$ reads
\begin{eqnarray}
(r^{D-4}f(r)c'(r))'-2k(D-3)r^{D-6}c(r)=\frac{p'(r)}{4\pi}.\label{30a}
\end{eqnarray}

To find the black hole solution, one may use any components of the equations of motion
Eq.~(\ref{3a}). While, these equations are influenced by charge and the $rr$-component reads
\begin{eqnarray}
-\frac{Q^2G}{2(D-2)\pi}r^{10-2D}
&=&[3\tilde{\alpha}_3r(f(r)-k)^2-2\tilde{\alpha}_2(f(r)-k)r^3+\tilde{\alpha}_1 r^5]f'(r)\nonumber\\
&+&(D-7)\tilde{\alpha}_3(f(r)-k)^3-(D-5)\tilde{\alpha}_2(f(r)-k)^2r^2\nonumber\\
&+&(D-3)\tilde{\alpha}_1(f(r)-k)r^4+(D-1)\tilde{\alpha}_0 r^6.\label{31a}
\end{eqnarray}
Setting $\tilde{\alpha}_0=-1/l^2$, we take the general charged solution $f(r)$ of D-dimensional slowly
rotating black hole in third order Lovelock gravity
\begin{eqnarray}
f(r)=k+\frac{r^2}{3\tilde{\alpha}_3}\Big[\tilde{\alpha}_2+\sqrt[3]{\sqrt{\gamma_*
+\kappa_*^2(r)}+\kappa_*(r)}-\sqrt[3]{\sqrt{\gamma_*+\kappa_*^2(r)}-\kappa_*(r)}\Big],\label{32a}
\end{eqnarray}
where
\begin{eqnarray}
\gamma_*=(3\tilde{\alpha}_1\tilde{\alpha}_3-{\tilde{\alpha}_2}^2)^3,\quad
\kappa_*(r)={\tilde{\alpha}_2}^3-\frac{9\tilde{\alpha}_1\tilde{\alpha}_2\tilde{\alpha}_3}{2}
-\frac{27{\tilde{\alpha}_3}^2}{2}[-1/l^2+\frac{m}{r^{D-1}}
-\frac{q^2}{r^{2D-4}}].\nonumber
\end{eqnarray}
We also introduce $f(r)=k-r^2\varphi_*$ with
\begin{eqnarray}
\varphi_*=-\frac{1}{3\tilde{\alpha}_3}\Big[\tilde{\alpha}_2+\sqrt[3]{\sqrt{\gamma_*+\kappa_*^2(r)}
+\kappa_*(r)}-\sqrt[3]{\sqrt{\gamma_*+\kappa_*^2(r)}-\kappa_*(r)}\Big].\label{33a}
\end{eqnarray}
The integration constant $M=\frac{(D-2)\Sigma_k}{16\pi G}m$  also is gravitational mass and the
charge $Q^2$ is expressed as $Q^2=\frac{2\pi(D-2)(D-3)}{G}q^2$.

In addition, there exist the off-diagonal $t\phi$-component of the equation of motion, which is
concerned with functions $p(r)$ and $c(r)$
\begin{eqnarray}
r^D(\tilde{\alpha}_1+2\tilde{\alpha}_2\varphi_*+3\tilde{\alpha}_3\varphi_*^2)p(r)'
=4GQ^2c(r)+C_3,\label{34a}
\end{eqnarray}
where $C_3$ is a constant.

We substitute metric Eq.~(\ref{5a}) into the action Eq.~(\ref{1a}).
Apparently, the forms of cosmological constant $\Lambda$, Einstein tensor $R$, Gauss-Bonnet
tensor ${\cal L}_2$  and third order Lovelock tensor ${\cal L}_3$ get no correction
from charge in the action and maintain the same form demonstrated in Eq.~(\ref{15a}).
As shown in Eq.~(\ref{29a}), there exist two non-vanishing $F_{r\phi}$ and $F_{\theta\phi}$
which are proportional to parameter $a$.
Discarding all terms involve $a^2$ and higher power, $F_{\mu\nu}F^{\mu\nu}$ in action Eq.~(\ref{1a})
reduces to $F_{tr}F^{tr}$ which is the same as the counterpart in static case.
Hence, $\varphi_*$ is determined by solving for the real roots of the following 3th
polynomial equation \cite{Myers:1988ze}
\begin{eqnarray}
\frac{1}{l^2}+\tilde{\alpha}_1\varphi_*+\tilde{\alpha}_2\varphi_*^2+\tilde{\alpha}_3\varphi_*^3
=\frac{m}{r^{D-1}}-\frac{q^2}{r^{2D-4}}.\label{35a}
\end{eqnarray}
Based on Eqs.~(\ref{30a})(\ref{34a})(\ref{35a}), we eventually find thses explicit solutions
for functions $p(r)$ and $c(r)$
\begin{eqnarray}
c(r)&=&-\frac{1}{4\pi(D-3)r^{D-3}}\nonumber\\
p(r)&=&-\varphi_*\nonumber\\
&=&\frac{1}{3\tilde{\alpha}_3}\Big[\tilde{\alpha}_2+\sqrt[3]{\sqrt{\gamma_*+\kappa_*^2(r)}
+\kappa_*(r)}-\sqrt[3]{\sqrt{\gamma_*+\kappa_*^2(r)}-\kappa_*(r)}\Big],\label{36a}
\end{eqnarray}
where $C_3$ is equal to $m(D-1)$.

In the rest of this section, let us explore some physical properties of charged black holes.
From Eq.~(\ref{32a}), the charged solutions get no corrections from the rotation up to linear
order of $a$, and the introduction of charged $Q$ does not alter asymptotic behavior of the metric.
Therefore, the expressions for the mass and angular momentum for two cases do not change. Another
particular characteristic of charged black hole is its gyromagnetic ratio. In general relativity,
one of the remarkable facts about a Kerr-Newman black hole in asymptotically flat spacetime
is that it can be assigned a gyromagnetic ratio $g$, just as an electron in the Dirac
theory \cite{Aliev:2007qi, Carter:1968rr}. For example, the gyromagnetic ratio $g$ of a
charged rotating black hole is $g=2$ in four-dimensional spacetime \cite{Aliev:2004ec}.
For slowly rotating third order Lovelock black holes, the magnetic dipole moment
is $\mu=Qa$. According to $J=\frac{2aM}{D-2}$, the gyromagnetic ratios is obtained
\begin{eqnarray}
g=\frac{2\mu M}{Q J}=D-2.\nonumber
\end{eqnarray}
It is clear that the value of $g$ is the same as the case in general
relativity \cite{Aliev:2007qi} and in Gauss-Bonnet gravity \cite{Kim:2007iw}, and it only
depends on the number of spacetime dimensions.

\section{Conclusion and Discussion\label{4s}}
\indent

Based on the non-rotating charged black hole solutions, we
successfully derived the slowly rotating (charged) black hole solutions by introducing
a small rotating parameter $a$ in third order Lovelock gravity. In the new metric,
we choose $g_{t\phi}=-ar^2p(r)h_{44}$
and discard any terms involving $a^2$ and higher powers, and then get the expression
for function $p(r)$, while the function $f(r)$ still keep the form of the static
solution. In charged case, the vector potential has an extra nonradial component
$A_{\phi}=-aQc(r)h_{44}$ due to the rotation of the black hole. Since the off-diagonal
component of the stress-tensor of electro-magnetic field was related to $c(r)$, the
equations for $p(r)$ and $c(r)$ become two non-homogeneous differential equations.
However, exact solutions for $c(r)$ and $p(r)$ have been separately
expressed as $c(r)=-\frac{1}{4\pi(D-3)r^{D-3}}$
and $p(r)=-\varphi_*$. In fact, it is still valid for general
$\tilde{\alpha}_i$. Up to the linear order of the rotating parameter $a$, the
expressions of the mass, temperature, and entropy for the black holes got no
correction from rotation in both uncharged and charged cases.

It is worth to point out that for third order Lovelock gravity,
its Lagrangian ${\cal L}_3$ involves eight terms constituted by
the Ricci and the Riemann curvature tensors. Moreover, the resulting field
equations, obtained after variation with respect to the metric tensor, have thirty-four
terms. Considering a higher order Lovelock term, for instance the quartic Lovelock tensor,
it involves twenty-five terms and each contains the product of four curvature terms.
A general expression of the corresponding field equations was obtained
in \cite{Briggs:1997ae}, while this work is very complicated. Therefore, taking into
account all the relevant terms of the Lovelock action, then obtaining slowly rotating
black hole solutions by solving the field equations for general space-times in high
dimensions, is a formidable task. Note that the exact static and spherically symmetric
black hole solutions of the Gauss-Bonnet gravity have been found by working directly
in the action \cite{Wheeler:1986, Cai:2003gr}.
Then, this simple method has been popularized in studying slowly rotating black holes in
third order Lovelock gravity \cite{Dehghani:2009zzb}, even higher order Lovelock
gravity \cite{Myers:1988ze}. However, the metric should be taken a
proper form. By using the same approach, the generalization of the present work may be
further simplified and is now under investigation.

{\bf Acknowledgment }

 This work has been supported by the National Natural Science Foundation
 of China under grant No. 10875060.

\end{document}